# Analytic Evaluation of two-electron Atomic Integrals involving Extended Hylleraas-CI functions with STO basis

B PADHY

(Retd.) Faculty Member
Department of Physics, Khalikote (Autonomous) College,
Berhampur-760001, Odisha
Email: bholanath.padhy@gmail.com

**Abstract :** Some typical overlap/potential energy integrals which occur in the use of extended Hylleraas-configuration interaction (E-Hy-CI) functions with Slater-type orbital (STO) basis for two-electron atomic structure calculations, have been evaluated analytically. The corresponding kinetic energy integrals have been simplified first by using formulas derived from Gauss′ divergence theorem in vector calculus, and then expressed in terms of overlap / potential energy matrix elements. Also closed-form expressions for such integrals which arise in the application of Hylleraas-CI functions, and CI functions have been obtained as special cases, and the calculated values are found to agree well with correct results published by other investigators.

**Keywords :** Exponentially correlated integrals, E-Hy-CI calculations, Two-electron atoms.

## 1. Introduction

It is well-accepted that electron-electron correlations in a multielectron atom are required to be included in the quantum mechanical calculations in order to obtain very accurate wave functions and energies for the atom. The methods that have been employed for this purpose are (i) Hylleraas (Hy) method [1,2], (ii) configuration-interaction (CI) method [3], (iii) Hylleraas-CI (Hy-CI) method [4] and (iv) extended Hy-CI (E-Hy-CI) method [5]. The E-Hy-CI method is an extension of Hy-CI method in the sense that exponential correlations are included in the former.

Details of the progressive development of Hy-CI method and its applications can be obtained by going through the recent papers [6-10] and the references therein. For knowledge about various calculations with E-Hy-CI method, the reader is advised to go through the recent papers [5,10-14].

In an earlier article [14], hereinafter referred as paper I, kinetic energy matrix elements for a two-electron atom have been expressed in terms overlap/potential energy integrals employing E-Hy-CI method. In this communication, which is rather termed as extension to paper I, a method of analytic evaluation of overlap/potential energy integrals involving E-Hy-CI functions has been outlined. As a result, it became possible to obtain values of several kinetic energy matrix elements easily.

## 2. Definition of two-electron overlap / potential energy integrals

Assuming the nucleus to be at rest, let $\vec{r}_1$ and $\vec{r}_2$ be the position vectors of the two electrons with respect to the nucleus. The distance between two electrons is denoted as $r_{12} = |\vec{r}_1 - \vec{r}_2|$. With spherical polar coordinates, the following set of unnormalised atomic Slater-type orbitals (STOs) is taken as the basis:

$$\varphi_a(j) = r_j^{n_a-1} e^{-\alpha_a r_j} Y_{l_a,m_a}(j); \quad \alpha_a > 0, \tag{1}$$

where $Y_{l_a,m_a}(j)$ is an orthonormal spherical harmonic with its argument being the angular coordinates of $\vec{r}_j$, and is defined in [4,8]. The subscript '$a$' signifies a particular STO. The most general type of overlap/potential energy integrals which arise in calculations with E-Hy-CI functions for a two-electron atomic system are of the form.

$$I^{EHCI} = \langle \varphi_a(1)\, \varphi_b(2) | r_{12}^\nu e^{-\beta r_{12}} | \varphi_d(1)\, \varphi_e(2) \rangle \tag{2}$$

with $\nu \geq -1$ and $\beta \geq 0$. Integrals $(I^{HCI})$ corresponding to Hy-CI functions can be obtained from Eq. (2), as a special case, by taking $\nu \geq 1$ and $\beta = 0$ simultaneously. Similarly if $\beta = 0$ and $\nu = 0$ simultaneously, one gets integrals $(I^{CI})$ corresponding to CI functions only.

### 2.1 Analytic evaluation of $I^{EHCI}$

From Eq. (2) it is observed that one may come across the product of two spherical harmonics of the same argument in the integrand. Such a product can be expanded as a series containing several terms with each term being a product of Clebsch-Gordan coefficients and an individual spherical harmonic of the same argument [15]. Hence, expressions for all the $I^{EHCI}$ integrals can be obtained by parametric differentiation of a basic integral given by

$$I_b^{EHCI} = \int d\vec{r}_1 \, r_1^{n_1-1} \, e^{-\omega_1 r_1} \, Y_{l_1,m_1}^*(\hat{r}_1) \, \mathcal{H}(\vec{r}_1, n_2, \omega_2, \beta, l_2, m_2), \qquad (3)$$

where

$$\mathcal{H}(\vec{r}_1, n_2, \omega_2, \beta, l_2, m_2) = \int d\vec{r}_2 \, r_2^{n_2-1} \, e^{-\omega_2 r_2} \, Y_{l_2,m_2}(\hat{r}_2) \, \frac{e^{-\beta r_{12}}}{r_{12}}, \qquad (4)$$

with the replacement of $\omega_1$ and $\omega_2$ by $\alpha_a, \alpha_b$, etc. wherever it is required.

It is to be pointed out here that a method was outlined by Calais and Lowdin [16] for evaluating such integrals as in Eq. (3), by employing the method of rotation of coordinate system, and subsequent transformation of spherical harmonics. In what follows, however, an alternative method of evaluation is employed which does not involve rotation of coordinate system. Accordingly, the above integral has been simplified in the Appendix-A to obtain

$$\mathcal{H}(\vec{r}_1, n_2, \omega_2, \beta, l_2, m_2) = Y_{l_2,m_2}(\theta_1, \varphi_1) \, G_{l_2}(r_1, n_2, \omega_2, \beta), \qquad (5)$$

where

$$G_{l_2}(r_1, n_2, \omega_2, \beta) = 2\pi \int_0^\infty dr_2 \, r_2^{n_2+1} e^{-\omega_2 r_2} \\
\times \int_{|r_1-r_2|}^{r_1+r_2} dr_{12} \, \frac{e^{-\beta r_{12}}}{r_1 r_2} \, P_{l_2}\left(\frac{r_1^2 + r_2^2 - r_{12}^2}{2 r_1 r_2}\right), \qquad (6)$$

for which a closed-form expression can be obtained, as well as for $\mathcal{H}$ through Eq. (5). Inserting Eq.(5) in Eq.(3) $I_b^{EHCI}$ becomes

$$I_b^{EHCI} = \delta_{l_1,l_2} \delta_{m_1,m_2} \int_0^\infty dr_1 \, r_1^{n_1+1} \, e^{-\omega_1 r_1} \, G_{l_2}(r_1, n_2, \omega_2, \beta), \qquad (7)$$

which can be easily evaluated analytically term by term using the formula

$$\int_0^\infty dx \, x^n \, e^{-ax} = \frac{\Gamma(n+1)}{a^{n+1}}. \qquad (8)$$

Alternatively, evaluation of $I_b^{EHCI}$ can be done by writing

$$I_b^{EHCI} = \delta_{l_1,l_2} \, \delta_{m_1,m_2} \times 2\pi \int_0^\infty dr_1 \, r_1^{n_1+1} e^{-\omega_1 r_1} \int_0^\infty dr_2 \, r_2^{n_2+1} e^{-\omega_2 r_2}$$

$$\times \int_{|r_1-r_2|}^{r_1+r_2} dr_{12} \frac{e^{-\beta r_{12}}}{r_1 r_2} P_{l_2}\left(\frac{r_1^2 + r_2^2 - r_{12}^2}{2 r_1 r_2}\right), \tag{9}$$

and following a method employed by Ruiz [17].

## 3. Reduction of kinetic energy integrals to integrable form

Various kinetic energy integrals, such as $K^{CI}, K_1^{HCI}, K_2^{HCI}, K_3^{HCI}$, $K^{EHCI}, K_1^{EHCI}, K_2^{EHCI}, K_3^{EHCI}$ have been defined and evaluated analytically in paper I. Hence, it is felt that there is no necessity of repeating those details here. However, simplification of two basic integrals will be done here by making use of Gauss' divergence theorem in vector calculus, and two related formulae, which have been derived in the Appendix-B. The following two integrals

$$K^{HCI}(v,v') = \left\langle r_{12}^v \, \varphi_a(1) \, \varphi_b(2) \left| -\tfrac{1}{2} \nabla_1^2 \right| \varphi_d(1) \, \varphi_e(2) \, r_{12}^{v'} \right\rangle; \quad v, v' \geq 0 \tag{10}$$

$$K^{EHCI}(w,w') = \left\langle \varphi_a(1) \, \varphi_b(2) \, e^{-w r_{12}} \left| -\tfrac{1}{2} \nabla_1^2 \right| \varphi_d(1) \, \varphi_e(2) \, e^{-w' r_{12}} \right\rangle; \quad w, w' \geq 0 \tag{11}$$

have been reduced to integrable form as given in the Appendix-C and Appendix-D, respectively. The simplified expressions are

$$K^{HCI}(v,v') = \frac{1}{2} \int d\vec{r}_2 \varphi_b^*(2) \varphi_e(2) \int d\vec{r}_1 \left[ v v' \varphi_a^*(1) \varphi_d(1) r_{12}^{v+v'-2} \right.$$
$$\left. - \frac{v'}{v+v'} r_{12}^{v+v'} \varphi_d(1) \vec{\nabla}_1^2 \varphi_a^*(1) - \frac{v}{v+v'} r_{12}^{v+v'} \varphi_a^*(1) \vec{\nabla}_1^2 \varphi_d(1) \right], \tag{12}$$

$$K^{EHCI}(w,w') = \frac{1}{2} \int d\vec{r}_2 \varphi_b^*(2) \varphi_e(2) \int d\vec{r}_1 \, e^{-(w+w') r_{12}}$$
$$\times \left[ \varphi_a^*(1) \varphi_d(1) w w' - \frac{w'}{w+w'} \varphi_d(1) \vec{\nabla}_1^2 \varphi_a^*(1) - \frac{w}{w+w'} \varphi_a^*(1) \vec{\nabla}_1^2 \varphi_d(1) \right]. \tag{13}$$

The integral in Eq. (10) has also been simplified by Ruiz [17].

The method followed here for simplification of the kinetic energy integrals has been known as Kolos-Roothan transformation method [18] in the literature.

Starting with Eqs. (10) and (12), definition of integrals and respective integrable expressions for $K^{CI}, K_1^{HCI}, K_2^{HCI}$ and $K_3^{HCI}$ can be obtained by choosing suitable values of $v$ and $v'$. By parametric differentiation of Eqs. (11) and (13) with respect to $w$ and/or $w'$, the definition of integrals $K_1^{EHCI}, K_2^{EHCI}$ and $K_3^{EHCI}$ as well as their integrable expressions as given in paper I can be reproduced.

*3.1 Analytic evaluation of kinetic energy integrals*

Expressing $\vec{\nabla}^2$ in spherical polar coordinates and employing some related formulae as given in the Appendix-B, it is straightforward to derive the following equations [8]:

$$-\frac{1}{2}\vec{\nabla}_1^2 \varphi_a^*(1) = \left[\frac{(l_a+n_a)(l_a-n_a+1)}{2r_1^2} + \frac{n_a \alpha_a}{r_1} - \frac{\alpha_a^2}{2}\right]\varphi_a^*(1), \quad (14)$$

$$-\frac{1}{2}\vec{\nabla}_1^2 \varphi_d(1) = \left[\frac{(l_d+n_d)(l_d-n_d+1)}{2r_1^2} + \frac{n_d \alpha_d}{r_1} - \frac{\alpha_d^2}{2}\right]\varphi_d(1). \quad (15)$$

Inserting Eqs. (14) and (15) in various equations containing integrable expressions, each of the kinetic energy integrals can be expressed as a combination of several overlap/potential energy integrals evaluated analytically in Appendix-A. Thus all the kinetic energy integrals are evaluated analytically. The closed-form expressions have been given in paper-I. Hence, it is not necessary to repeat the expressions here.

The values of several kinetic energy integrals have been calculated and observed to be in close agreement with those reported by Ruiz [17] and by Harris [9].

## 4. Conclusion

The matrix elements for the kinetic energy integrals for a two-electron atom corresponding to CI, Hy-CI and E-Hy-CI functions could be expressed in terms of the matrix elements for the overlap/potential energy integrals for which closed-form expressions have been obtained here. Calculated values of various such integrals will be displayed in the form of tables and compared with those published by other investigators [9,17] in a future publication.

**Acknowledgements**

The author is highly indebted to Dr. J.S. Sims and Dr. M.B. Ruiz for periodic discussions relating to this work through exchange of e-mails. Also, I am very much grateful to Prof. N. Barik for several useful suggestions, and to Dr. P.K. Panda for help during computation and for critically going through the manuscript.

**Appendix A: Equation (5) established**

In this Appendix, a closed-form expression for the following integral $\mathcal{H}$ will be derived:

$$\mathcal{H}(\vec{r}_1, n_2, \omega_2, \beta, l_2, m_2) = \int d\vec{r}_2 \, r_2^{n_2-1} e^{-\omega_2 r_2} Y_{l_2,m_2}^{(\hat{r}_2)} \frac{e^{-\beta r_{12}}}{r_{12}}. \tag{A1}$$

To evaluate this integral spherical polar coordinates are employed, in which the volume element $d\vec{r}_2$ is replaced by $r_2^2 \, dr_2 \sin\theta_2 \, d\theta_2 \, d\varphi_2$, and hence Eq. (A1) is written as

$$\mathcal{H} = \int_0^\infty dr_2 r_2^{n_2+1} e^{-\omega_2 r_2} \int_0^\pi \int_0^{2\pi} \sin\theta_2 \, d\theta_2 \, d\varphi_2 \, Y_{l_2,m_2}(\theta_2, \varphi_2) \frac{e^{-\beta r_{12}}}{r_{12}}. \tag{A2}$$

Here

$$r_{12} = (r_1^2 + r_2^2 - 2r_1 r_2 \cos\theta_{12})^{1/2}; \theta_{12} = \cos^{-1}(\hat{r}_1 \cdot \hat{r}_2). \tag{A3}$$

It is clear that $e^{-\beta r_{12}}/r_{12}$ is a function of $r_1, r_2, \beta$ and $\cos\theta_{12}$. It is also known that the Legendre polynomials $P_L(\cos\theta_{12})$ of degree $L = 0,1,2,\ldots$, etc. constitute a complete set of orthogonal polynomials satisfying the following orthogonality condition :

$$\int_0^\pi P_L(\cos\theta_{12}) P_{L'}(\cos\theta_{12}) \sin\theta_{12} \, d\theta_{12} = \frac{2}{2L+1} \delta_{L,L'}. \tag{A4}$$

Therefore, $e^{-\beta r_{12}}/r_{12}$ can be expanded in the form of the following series :

$$\frac{e^{-\beta r_{12}}}{r_{12}} = \sum_{L=0}^{\infty} a_L(r_1, r_2, \beta)\, P_L(\cos\theta_{12}), \qquad (A5)$$

where $a_L s$ are the coefficients in the expansion to be determined.

Multiplying both sides of Eq. (A5) by $P_{L'}(\cos\theta_{12})\sin\theta_{12}\, d\theta_{12}$, then integrating over $\theta_{12}$ ($0 \leq \theta_{12} \leq \pi$), and employing Eq. (A4), it is easy to obtain

$$a_{L'}(r_1, r_2, \beta) = \frac{2L'+1}{2} \int_0^\pi \frac{e^{-\beta r_{12}}}{r_{12}} P_{L'}(\cos\theta_{12}) \sin\theta_{12} d_{12}. \qquad (A6)$$

Inserting Eq. (A5) in Eq. (A2), then employing the spherical harmonic addition theorem

$$P_L(\cos\theta_{12}) = \frac{4\pi}{2L+1} \sum_{M=-L}^{L} Y_{L,M}^*(\theta_1, \varphi_1) Y_{L,M}(\theta_2, \varphi_2), \qquad (A7)$$

and making use of the orthonormality property of the spherical harmonics, the angular integration in Eq. (A2) is done to get

$$\mathcal{H}(\vec{r}_1, n_2, \omega_2, \beta, l_2, m_2) = Y_{l_2, m_2}(\theta_1, \varphi_1)\, G_{l_2}(r_1, n_2, \omega_2, \beta), \qquad (A8)$$

where

$$G_{l_2}(r_1, n_2, \omega_2, \beta) = \frac{4\pi}{2l_2+1} \int_0^\infty dr_2\, r_2^{n_2+1} e^{-\omega_2 r_2}\, a_{l_2}(r_1, r_2, \beta). \qquad (A9)$$

Inserting Eq. (A6) in Eq. (A9), one obtains

$$G_{l_2} = 2\pi \int_0^\infty dr_2\, r_2^{n_2+1} e^{-\omega_2 r_2} \int_0^\pi \frac{e^{-\beta r_{12}}}{r_{12}} P_{l_2}(\cos\theta_{12}) \sin\theta_{12}\, d\theta_{12}, \qquad (A10)$$

which indicates that $G_{l_2}$ depends on $l_2$ but not on $m_2$, through the presence of the Legendre polynomial in the integrand. To obtain a closed-form expression for $G_{l_2}$, and hence for $\mathcal{H}$ through Eq. (A8), change of variable from $\theta_{12}$ to $r_{12}$ is made using Eq. (A3). Thus

$$\cos\theta_{12} = \frac{r_1^2 + r_2^2 - r_{12}^2}{2r_1 r_2} \qquad (A11)$$

and

$$\sin\theta_{12}\, d\theta_{12} \to \frac{r_{12}\, dr_{12}}{r_1 r_2}. \qquad (A12)$$

Incorporating these changes in Eq. (A10), the following integrable expression for $G_{l_2}$ is obtained:

$$G_{l_2}(r_1, n_2, \omega_2, \beta) = 2\pi \int_0^\infty dr_2\, r_2^{n_2+1} e^{-\omega_2 r_2}$$

$$\times \int_{|r_1-r_2|}^{r_1+r_2} d r_{12}\, \frac{e^{-\beta r_{12}}}{r_1 r_2} P_{l_2}\!\left(\frac{r_1^2 + r_2^2 - r_{12}^2}{2 r_1 r_2}\right). \qquad (A13)$$

For a particular value of $l_2$, the Legender polynomial $P_{l_2}(\cos\theta_{12})$ is replaced by a function of $r_1, r_2$ and $r_{12}$ using Eq. (A11). Then integration in Eq. (A13) is performed easily to obtain a closed-form expression for $G_{l_2}$. Thus the integral $\mathcal{H}$ as defined in Eq. (A1) is analytically evaluated using Eq. (A8).

## Appendix B: Some preliminaries relating to $\vec{\nabla}$

The operator nabla, also called del, and atled (inverse of delta), is a vector differential operator defined in Cartesian coordinates by the relation

$$\vec{\nabla} = \hat{i}\frac{\partial}{\partial x} + \hat{j}\frac{\partial}{\partial y} + \hat{k}\frac{\partial}{\partial z}, \qquad (B1)$$

and is widely used in vector calculus. Only some preliminaries, which are useful for the present investigation, are outlined below.

Let $\psi$ and $\varphi$ be two scalar point functions, and $\vec{F}$ be a vector point function, which are, in general, complex well-behaved functions vanishing at infinity. The Gauss′ divergence theorem in vector calculus is mathematically stated as

$$\iiint_V (\vec{\nabla} \cdot \vec{F})\, dv = \iint_S \vec{F} \cdot d\vec{s} \qquad (B2)$$

with the condition that V is the volume enclosed by the closed surface S. If the volume integral on the left hand side of Eq. (B2) is evaluated over the entire space, then the surface integral on the right hand side of Eq.(B2) vanishes because $\vec{F}$ vanishes at each point on the surface enclosing the infinite volume as per supposition. This conclusion will be used in the following paragraph.

Using the vector identity

$$\vec{\nabla} \cdot (\psi \vec{F}) = (\vec{\nabla} \psi) \cdot \vec{F} + \psi (\vec{\nabla} \cdot \vec{F}), \qquad (B3)$$

and then integrating both sides of Eq.(B3) over the whole space, the integral on the left hand side is reduced to a surface integral as per Eq. (B2), and vanishes at infinity since the product $\psi \vec{F} \to 0$ at each point on the infinite surface as per supposition. Hence the following relation is established:

$$\iiint (\vec{\nabla} \psi) \cdot \vec{F} \, dv = -\iiint \psi (\vec{\nabla} \cdot \vec{F}) \, dv, \qquad (B4)$$

with the integration to be done over the entire space. Similarly, integrating both sides of the following identity

$$\vec{\nabla} \cdot (\psi^* \vec{\nabla} \varphi) = (\vec{\nabla} \psi^*) \cdot (\vec{\nabla} \varphi) + \psi^* (\vec{\nabla}^2 \varphi), \qquad (B5)$$

over the whole space, then employing Eq. (B2) and noting that $\psi^* \to 0$ at each point on the infinite surface, the following equation is obtained:

$$\iiint \psi^* (\vec{\nabla}^2 \varphi) \, dv = -\iiint (\vec{\nabla} \psi^*) \cdot (\vec{\nabla} \varphi) \, dv. \qquad (B6)$$

In particular, if $\psi = \varphi,$ then Eq.(B6) becomes

$$\iiint \varphi^* (\vec{\nabla}^2 \varphi) \, dv = -\iiint |\vec{\nabla} \varphi|^2 \, dv. \qquad (B7)$$

Another property of the nabla operator is that it is antihermitian, which follows from the definition of adjoint of an operator, and a Hermitian operator. Since the linear momentum operator $\hat{p} = -i\hbar \vec{\nabla}$ is Hermitian, it follows that $\vec{\nabla}^+ = -\vec{\nabla}$, that is, the nabla operator is antihermitian.

In spherical polar coordinates $(r, \theta, \varphi)$

$$\vec{\nabla} = \hat{r} \frac{\partial}{\partial r} + \hat{\theta} \frac{1}{r} \frac{\partial}{\partial \theta} + \hat{\varphi} \frac{1}{r \sin \theta} \frac{\partial}{\partial \varphi}, \qquad (B8)$$

and

$$\nabla^2 = \frac{\partial^2}{\partial r^2} + \frac{2}{r}\frac{\partial}{\partial r} - \frac{\hat{L}^2}{\hbar^2 r^2}, \tag{B9}$$

where $\hat{L}^2$ is the operator corresponding to the square of the angular momentum and is given by

$$\hat{L}^2 \equiv -\hbar^2\left[\frac{1}{\sin\theta}\frac{\partial}{\partial\theta}\left(\sin\theta\frac{\partial}{\partial\theta}\right) + \frac{1}{\sin^2\theta}\frac{\partial^2}{\partial\varphi^2}\right]. \tag{B10}$$

The following are few established equations:

$$\hat{L}^2 Y_l^m(\theta,\varphi) = l(l+1)\hbar^2 Y_l^m(\theta,\varphi), l = 0,1,2, \text{etc.}; \tag{B11}$$

$$\hat{L}_z Y_l^m(\theta,\varphi) = m\hbar Y_l^m(\theta,\varphi), m = l, l-1, ..., 0, -1, -2, ..., -l; \tag{B12}$$

$$\vec{\nabla}^2 Y_l^m(\theta,\varphi) = -l(l+1) Y_l^m(\theta,\varphi)/r^2; \tag{B13}$$

$$\vec{\nabla} f(r) = f'(r)\vec{\nabla} r = f'(r)\hat{r};$$

(B14)

$$\vec{\nabla}_1 r_{12} = \frac{\vec{r}_1 - \vec{r}_2}{r_{12}} = \hat{r}_{12}; \tag{B15}$$

$$\vec{\nabla}_2 r_{12} = \frac{\vec{r}_2 - \vec{r}_1}{r_{12}} = -\hat{r}_{12}. \tag{B16}$$

**Appendix-C : Equation (12) established**

The general kinetic energy integral to be evaluated in Hy-CI calculations, as defined by Eq. (10) in the text, is

$$K^{HCI}(\nu,\nu') = \langle r_{12}^\nu \varphi_a(1)\varphi_b(2) \left| -\tfrac{1}{2}\vec{\nabla}_1^2 \right| \varphi_d(1)\varphi_e(2) r_{12}^{\nu'}\rangle, \tag{C1}$$

where $\nu,\nu' \geq 0$ (integers). The above equation can be rewritten as

$$K^{HCI}(\nu,\nu') = \int d\vec{r}_2\, \varphi_b^*(2)\varphi_e(2)\, K^a(\nu,\nu',\vec{r}_2), \tag{C2}$$

where

$$K^a(\nu,\nu',\vec{r}_2) = \langle r_{12}^\nu \varphi_a(1) | -\tfrac{1}{2}\vec{\nabla}_1^2 | \varphi_d(1) r_{12}^{\nu'} \rangle. \tag{C3}$$

Since $\vec{\nabla}_1^2 = \vec{\nabla}_1 \cdot \vec{\nabla}_1$ and $\vec{\nabla}_1^+ = -\vec{\nabla}_1$, Eq. (B6) is employed to get

$$K^a(\nu,\nu',\vec{r}_2) = \tfrac{1}{2}\int d\vec{r}_1 \{\vec{\nabla}_1(r_{12}^\nu \varphi_a^*(1))\} \cdot \{\vec{\nabla}_1(\varphi_d(1) r_{12}^{\nu'})\}. \tag{C4}$$

Expanding the gradient of the scalar point functions within each pair of curly brackets and then taking the scalar product, one obtains

$$K^a(\nu,\nu',\vec{r}_2) = \tfrac{1}{2}\int d\vec{r}_1 \Big[ r_{12}^{\nu+\nu'}(\vec{\nabla}_1\varphi_a^*)\cdot(\vec{\nabla}_1\varphi_d) + \nu\nu'\varphi_a^*\varphi_d \, r_{12}^{\nu+\nu'-2}$$
$$+ (r_{12}^{\nu+\nu'-1}\hat{r}_{12})\cdot\{\nu'\varphi_d(\vec{\nabla}_1\varphi_a^*) + \nu\varphi_a^*(\vec{\nabla}_1\varphi_d)\} \Big], \tag{C5}$$

where $\hat{r}_{12} = (\vec{r}_1 - \vec{r}_2)/r_{12}$. Also $\varphi_a^*(1)$ and $\varphi_d(1)$ are replaced by $\varphi_a^*$ and $\varphi_d$, respectively, for convenience. Making use of the relation

$$r_{12}^{\nu+\nu'-1}\hat{r}_{12} = \vec{\nabla}_1\left(\frac{r_{12}^{\nu+\nu'}}{\nu+\nu'}\right), \tag{C6}$$

the integral corresponding to the third term of the integrand in Eq. (C5) can be simplified as per Eq. (B4). Consequently, expanding the divergence of the vector point function in the resulting integrand and adding with the first two terms of the integrand in Eq. (C5), the following expression for $K^a$ is obtained.

$$K^a(\nu,\nu',\vec{r}_2) = \tfrac{1}{2}\int d\vec{r}_1 \Big[ \nu\nu'\varphi_a^*(1)\varphi_d(1) r_{12}^{\nu+\nu'-2}$$
$$-\frac{\nu'}{\nu+\nu'} r_{12}^{\nu+\nu'}\varphi_d(\vec{\nabla}_1^2\varphi_a^*) - \frac{\nu}{\nu+\nu'} r_{12}^{\nu+\nu'}\varphi_a^*(\vec{\nabla}_1^2\varphi_d) \Big]. \tag{C7}$$

Inserting Eq. (C7) in Eq. (C2), the relation given in Eq.(12) in the text is established.

## Appendix D : Equation (13) established

As per Eq. (11) in the text, the integral $K^{EHCI}(w,w')$ is given by

$$K^{EHCI}(w,w') = \langle \varphi_a(1)\varphi_b(2) e^{-w r_{12}} | -\tfrac{1}{2}\vec{\nabla}_1^2 | \varphi_d(1)\varphi_e(2) e^{-w' r_{12}} \rangle, \tag{D1}$$

where $w$ and $w'$ are the exponential parameters with $w, w' \geq 0$. The integral can be recast as

$$K^{EHCI} = \int d\vec{r}_2\, \varphi_b^*(2)\, \varphi_e(2)\, K_a^{EHCI}(w, w', \vec{r}_2),\qquad (D2)$$

where

$$K_a^{EHCI} = \langle \varphi_a(1)\, e^{-w r_{12}} \,\big|\, -\tfrac{1}{2} \vec{\nabla}_1^2 \,\big|\, \varphi_d(1)\, e^{-w' r_{12}} \rangle. \qquad (D3)$$

Since $\vec{\nabla}_1^2 = \vec{\nabla}_1 \cdot \vec{\nabla}_1$ and $\vec{\nabla}_1^+ = -\vec{\nabla}_1$, Eq. (D3) becomes

$$K_a^{EHCI} = \tfrac{1}{2} \int d\vec{r}_1 \left\{\vec{\nabla}_1(\varphi_a^*(1)\, e^{-w r_{12}})\right\} \cdot \left\{\vec{\nabla}_1(\varphi_d(1)\, e^{-w' r_{12}})\right\}. \qquad (D4)$$

Expanding the gradient of the scalar point function within each pair of the curly brackets, the scalar product of the resulting vector point functions is obtained. Taking note that $\vec{\nabla}_1 r_{12} = \hat{r}_{12}$, Eq. (D4) is simplified to get

$$\begin{aligned}
K_a^{EHCI} = \tfrac{1}{2} \int d\vec{r}_1\, e^{-(w+w') r_{12}} \Big[&(\vec{\nabla}_1 \varphi_a^*(1)) \cdot (\vec{\nabla}_1 \varphi_d(1)) + \varphi_a^*(1)\, \varphi_d(1)\, w w' \\
& - \hat{r}_{12} \cdot \left\{w' \varphi_d(1)(\vec{\nabla}_1 \varphi_a^*(1)) + w \varphi_a^*(1)(\vec{\nabla}_1 \varphi_d(1))\right\}\Big]. \quad (D5)
\end{aligned}$$

Since

$$e^{-(w+w') r_{12}}\, \hat{r}_{12} = \vec{\nabla}_1 \left\{ \frac{e^{-(w+w') r_{12}}}{-(w+w')} \right\}, \qquad (D6)$$

the third term of the integral, denoted as $T_3$, in Eq. (D5) becomes

$$T_3 = -\tfrac{1}{2} \int d\vec{r}_1 \left[ \vec{\nabla}_1 \left\{ \frac{e^{-(w+w') r_{12}}}{-(w+w')} \right\} \right] \cdot \left\{ w' \varphi_d(1)\, \vec{\nabla}_1 \varphi_a^*(1) + w \varphi_a^*(1)\, \vec{\nabla}_1 \varphi_d(1) \right\}, \qquad (D7)$$

which is further simplified by using Eq. (B4). Thus

$$T_3 = -\frac{1}{2(w+w')} \int d\vec{r}_1\, e^{-(w+w') r_{12}}\, \vec{\nabla}_1 \cdot \left\{ w' \varphi_d(1)\, \vec{\nabla}_1 \varphi_a^*(1) + w \varphi_a^*(1)\, \vec{\nabla}_1 \varphi_d(1) \right\}. \qquad (D8)$$

Expanding the divergence of the sum of the two vector point functions within the curly brackets in the integrand in Eq. (D8), and then inserting the resulting expression in Eq. (D5), some cancellations are made to obtain

$$K_a^{EHCI} = \frac{1}{2} \int d\vec{r}_1 \, e^{-(w+w')r_{12}} \left[ \varphi_a^*(1)\varphi_d(1) \, ww' \right.$$

$$\left. - \frac{w'}{w+w'} \, \varphi_d(1) \overleftarrow{\nabla}_1^2 \varphi_a^*(1) - \frac{w}{w+w'} \varphi_a^*(1) \vec{\nabla}_1^2 \varphi_d(1) \right]. \quad \text{(D9)}$$

Inserting Eq. (D9) in Eq. (D2), one gets

$$K^{EHCI}(w,w') = \frac{1}{2} \int d\vec{r}_2 \, \varphi_b^*(2) \, \varphi_e(2) \int d\vec{r}_1 \, e^{-(w+w')r_{12}}$$

$$\times \left[ \varphi_a^*(1)\varphi_d(1) \, ww' - \frac{w'}{w+w'} \varphi_d(1) \, \vec{\nabla}_1^2 \varphi_a^*(1) - \frac{w}{w+w'} \varphi_a^*(1) \vec{\nabla}_1^2 \varphi_d(1) \right], \quad \text{(D10)}$$

which is the required Eq. (13) in the text.

**References**

[1] EA Hylleraas, *Adv. Quantum Chem.* **1**, 1(1964)

[2] FW King, *Recent Advances in Computational Chemistry*, *Molecular Integrals over Slater Orbitals*, edited by T Ozdogan and MB Ruiz (Transworld, Kerala, India), pp. 39-84 (2008)

[3] JN Silverman and GH Brigman, *Rev. Mod. Phys.* **39**, 228 (1967)

[4] JS Sims and SA Hagstrom, *J. Chem. Phys.* 5**5**, 4699 (1971)

[5] C Wang, P Mei, Y Kurokawa, H Nakashima and H Nakatsuji, *Phys. Rev.* A **85**, 042512 (2012)

[6] JS Sims and SA Hagstrom, *J. Phys. B*: *At. Mol. Opt. Phys.* **48**, 175003 (2015)

[7] MB Ruiz, *J. Math. Chem.* **54**, 1083 (2016)

[8] FE Harris, *J. Chem. Phys.* **144**, 204110 (2016); *erratum* **145**, 129901 (2016)

[9] FE Harris, *Mol. Phys.* **115** (17-18), 2048 (2017)

[10] FE Harris, *Adv. Quantum Chem.* **76**, 187 (2017)

[11] B Padhy, *Asian J. Spectrosc.* (special Issue) pp. 157-162 (2012); arXiv: 1609.00269

[12] FW King, *J. Phys. B: At. Mol. Opt. Phys.* **49**, 105001 (2016)

[13] B Padhy, arXiv:1701.05701

[14] B Padhy, arXiv: 1801.10322; *Orissa J. Phys.* **25 (1)**, 9 (2018)

[15] ME Rose, *Elementary Theory of Angular Momentum*, Dover Publications, INC. New York (1957), pp.61.

[16] JL Calais and PO Lowdin, *J. Mol. Spectrosc.* **8**, 203 (1962)

[17] MB Ruiz, *J. Math. Chem.* **49**, 2457 (2011)

[18] W Kolos and CCJ Roothan, *Rev. Mod. Phys.* **32**, 219 (1960)